\documentclass[a4paper,12pt]{article}

\usepackage[dvips]{graphicx}
\usepackage{epsfig}

\begin{document}

\author{C. Barrab\`es\thanks{E-mail : barrabes@lmpt.univ-tours.fr}\\
\small Laboratoire de Math\'ematiques et Physique Th\'eorique,\\
\small CNRS/UMR 6083, Universit\'e F. Rabelais, 37200 TOURS,
France
\\\small and \\P. A. Hogan\thanks{E-mail : peter.hogan@ucd.ie}\\
\small School of Physics,\\ \small University College Dublin,
Belfield, Dublin 4, Ireland}

\title{The Bell--Szekeres Solution and Related Solutions of the Einstein--Maxwell Equations}
\date{}
\maketitle

\begin{abstract}
A novel technique for solving some head--on collisions of plane
homogeneous light--like signals in Einstein--Maxwell theory is
described. The technique is a by--product of a re--examination of
the fundamental Bell--Szekeres solution in this field of study.
Extensions of the Bell--Szekeres collision problem to include
light--like shells and gravitational waves are described and a
family of solutions having geometrical and topological properties
in common with the Bell--Szekeres solution is derived.

\end{abstract}
\thispagestyle{empty}
\newpage

\section{Introduction}\indent
The simplest collision space--time known involving homogeneous
plane electromagnetic waves in Einstein--Maxwell theory is
arguably the head--on collision of electromagnetic shock waves
having a step function profile given by Bell and Szekeres
\cite{BS}. The Bell--Szekeres solutions satisfy the vacuum
Einstein--Maxwell field equations while the resulting space--times
admit a pair of space--like, hypersurface--orthogonal, commuting
Killing vector fields. The line--element of the region of the
Bell--Szekeres space--time following the collision of the shock
waves turned out to be identical to the line--element found
independently by Bertotti \cite{Ber} and by Robinson \cite{R},
whose motivation was purely geometrical. Global properties of the
Bell--Szekeres solutions were subsequently described by Clarke and
Hayward \cite{CH}.

The physical interpretation of the Bell--Szekeres solutions is
that they describe the electromagnetic and gravitational fields
before and after the head--on collision of two linearly polarized,
homogeneous, plane fronted electromagnetic shock waves which (a)
have a step function profile and (b) which do not interact before
collision.

By analogy with the Weyl static, axially symmetric space--times or
more generally the stationary, axially symmetric space--times,
which admit two commuting Killing vector fields (one space--like
and one time--like) which can be described in the formulation of
Ernst \cite{E1} \cite{E2}, the space--times resulting from the
head--on collision of plane fronted, homogeneous, light--like
signals can also be described in an Ernst formalism (see
Chandrasekhar \cite{C}, Chandrasekhar and Ferrari \cite{CF} and
Chandrasekhar and Xanthopoulos \cite{CX}).

In this paper we consider the problem of finding the space--time
after the head--on collision of two homogeneous, plane fronted,
light--like signals, each of which incorporates an electromagnetic
shock wave of the Bell--Szekeres type, an impulsive gravitational
wave and a light--like shell of matter. The solution to this
problem is in general unknown and, among other special cases, it
must include the Bell--Szekeres solutions. We present in this
paper new solutions of the Einstein--Maxwell field equations which
constitute a subclass of this general problem for which there
exist algebraic relations between the parameters describing the
different physical components of the incoming light--like signals.
Our family of solutions contains the Bell--Szekeres solutions as a
special case and from a geometrical and topological point of view
are shown to be closely related to the Bell--Szekeres
space--times. We make use of a technique for solving the field
equations which we were led to while re--constructing the
Bell--Szekeres space--times. No derivation of the solutions is
described in the original paper by Bell and Szekeres. Our
technique is both simple and new and has the possibility of being
used to find further new solutions of these field equations.

The outline of the paper is as follows: In section 2 we describe a
novel technique for finding collision solutions of the Maxwell and
Einstein--Maxwell vacuum field equations and illustrate it by
giving a derivation of the Bell--Szekeres solution. In section 3
we state the general collision problem in which the
electromagnetic shock waves of Bell--Szekeres are accompanied by
light--like shells of matter and impulsive gravitational waves. In
commenting on the general problem we describe a new collision
solution which solves a subclass of the general problem. A
subclass of the general collision problem described in section 3
which includes the Bell--Szekeres solution as a special case is
derived in detail in section 4 making use of the technique given
in section 2. Consequences of conformal flatness are worked out in
the present context in section 5 and establish the uniqueness of
the collision problem solved in section 4. A topological property
of the family of solutions obtained in section 4 is discussed in
section 6.

\setcounter{equation}{0}
\section{The Bell--Szekeres Solution Revisited}\indent
As in all collision problems involving the head--on collision of
homogeneous plane light--like signals, going back to the
pioneering work of Szekeres \cite{S1}\cite{S2}, Khan and Penrose
\cite{KP} and Bell and Szekeres \cite{BS} we start with the
Rosen--Szekeres form of line--element:
\begin{equation}\label{1.1}
ds^2=-{\it e}^{-U}\,({\it e}^{V}\,dx^2+{\it e}^{-V}\,dy^2)+2\,{\it
e}^{-M}\,du\,dv\ ,\end{equation}where the functions $U, V, M$
depend on the coordinates $u, v$ only. The Maxwell field is
described in Newman--Penrose notation by two real--valued
functions $\phi _0$ and $\phi _2$ (depending only on $u, v$)
satisfying Maxwell's vacuum field equations,
\begin{eqnarray}\label{1.2}
\frac{\partial\phi _2}{\partial v}&=&\frac{1}{2}U_v\,\phi
_2-\frac{1}{2}V_u\,\phi _0\ ,\\
\frac{\partial\phi _0}{\partial u}&=&\frac{1}{2}U_u\,\phi
_0-\frac{1}{2}V_v\,\phi _2\ .\end{eqnarray}Here and throughout
subscripts denote partial derivatives. The functions appearing in
the line--element (\ref{1.1}) satisfy the Einstein--Maxwell vacuum
field equations:
\begin{eqnarray}\label{1.3}
U_{uv}&=&U_u\,U_v\ ,\\
2\,U_{uu}&=&U_u^2+V_u^2-2\,U_u\,M_u+4\,\phi _2^2\ ,\\
2\,U_{vv}&=&U_v^2+V_v^2-2\,U_v\,M_v+4\,\phi _0^2\ ,\\
2\,V_{uv}&=&U_u\,V_v+U_v\,V_u+4\,\phi _0\,\phi _2\ ,\\
2\,M_{uv}&=&V_u\,V_v-U_u\,U_v\ .\end{eqnarray}It is well--known
that the first of these is solved in general by
\begin{equation}\label{1.4}
{\it e}^{-U}=f(u)+g(v)\ ,\end{equation}and the functions $f$ and
$g$ are immediately determined by the initial (boundary)
conditions. As a final preliminary we note that the
Newman--Penrose components of the Weyl conformal curvature tensor
calculated with the metric given via the line--element (\ref{1.1})
are
\begin{eqnarray}\label{1.4'}
\Psi _0&=&-\frac{1}{2}\left (V_{vv}-U_v\,V_v+M_v\,V_v\right )\ ,\\
\Psi _1&=&0\ ,\\
\Psi _2&=&\frac{1}{4}\left (V_u\,V_v-U_u\,U_v\right )\ ,\\
\Psi _3&=&0\ ,\\
\Psi _4&=&-\frac{1}{2}\left (V_{uu}-U_u\,V_u+M_u\,V_u\right )\
.\end{eqnarray}

We describe here a simple technique to solve the field equations
(2.2)--(2.8) in special circumstances. Since the technique arose
in our re--examination of the Bell--Szekeres solution we will
illustrate it by using it to give a derivation of the
Bell--Szekeres solution. We begin by re--writing (2.2) in the form
\begin{equation}\label{1.5}
\frac{\partial}{\partial v}(\log\phi
_2)=\frac{1}{2}U_v-\frac{1}{2}V_u\,\frac{\phi _0}{\phi _2}\
,\end{equation}and re--writing (2.3) in the form
\begin{equation}\label{1.6}
\frac{\partial}{\partial u}(\log\phi
_0)=\frac{1}{2}U_u-\frac{1}{2}V_v\,\frac{\phi _2}{\phi _0}\
.\end{equation}From these we deduce that
\begin{equation}\label{1.7}
2\,\frac{\partial ^2}{\partial u\partial v}(\log\frac{\phi
_2}{\phi _0})=\frac{\partial}{\partial v}\left (V_v\,\frac{\phi
_2}{\phi _0}\right )-\frac{\partial}{\partial u}\left
(V_u\,\frac{\phi _0}{\phi _2}\right )\ .\end{equation}

We note that all of the equations given so far are invariant under
the transformations $u\rightarrow\bar u=\bar u(u)$ and
$v\rightarrow\bar v=\bar v(v)$. Under these transformations the
functions $\phi _0$, $\phi _2$ and $M$ transform as
\begin{equation}\label{1.8}
\phi _0\rightarrow\bar\phi _0\ ,\qquad\phi _2\rightarrow\bar\phi
_2\ ,\qquad M\rightarrow\bar M\ ,\end{equation}with
\begin{equation}\label{1.9}
\phi _0=\frac{d\bar v}{dv}\,\bar\phi _0\ ,\qquad \phi
_2=\frac{d\bar u}{du}\,\bar\phi _2\ ,\qquad {\it e}^{\bar M}={\it
e}^M\,\frac{d\bar u}{du}\,\frac{d\bar v}{dv}\ .\end{equation}We
will be interested in seeking solutions of (\ref{1.7}) for which
\begin{equation}\label{1.10}
\frac{\phi _2}{\phi _0}=\frac{A(u)}{B(v)}\ ,\end{equation}for some
functions $A(u)$ and $B(v)$. In this case we can use (\ref{1.8})
and (\ref{1.9}) to choose a frame $(\bar u, \bar v)$ for which
\begin{equation}\label{1.11}
\bar\phi _0=\bar\phi _2\ .\end{equation}After the head--on
collision of the electromagnetic waves the functions $\phi _0$ and
$\phi _2$ describe back--scattered electromagnetic waves and
(\ref{1.11}) implies that there exists a frame of reference in
which the energy densities of these back--scattered waves are
equal. In this frame the equation (\ref{1.7}) becomes a wave
equation for $V$:
\begin{equation}\label{1.12}
V_{\bar u\bar u}=V_{\bar v\bar v}\ .\end{equation}If when $\bar
v=0$ we have the initial data, $V=P(\bar u)\ ,\ V_{\bar v}=Q(\bar
u)$ then $V$ is given for $\bar u>0\ ,\ \bar v >0$ by the
d'Alembert formula:
\begin{equation}\label{1.13}
V(\bar u, \bar v)=\frac{1}{2}\left\{P(\bar u+\bar v)+P(\bar u-\bar
v)\right\}+\frac{1}{2}\int_{\bar u-\bar v}^{\bar u+\bar v}Q(\xi
)\,d\xi\ .\end{equation}

To illustrate the method in the previous paragraph we take the
Bell--Szekeres problem. This consists of looking for the
space--time following the head--on collision of two
electromagnetic shock waves each having a step function profile.
In terms of the coordinates introduced at the beginning of this
section $\{x, y, u, v\}$ and the functions $U, V, M, \phi _0, \phi
_2$ this problem is expressed mathematically by requiring a
solution of the Einstein--Maxwell field equations (2.2)--(2.8)
with the initial conditions: for $v=0,\ u>0$ we require
\begin{equation}\label{1.14}
{\it e}^{U}=1+a^2u^2\ ,\ V=0\ ,\ {\it e}^M=1+a^2u^2\ ,\ \phi
_2=\frac{a}{1+a^2u^2}\ ,\end{equation}where $a$ is a real
constant, and for $u=0,\ v>0$ we require
\begin{equation}\label{1.15}
{\it e}^{U}=1+b^2v^2\ ,\ V=0\ ,\ {\it e}^M=1+b^2v^2\ ,\ \phi
_0=\frac{b}{1+b^2v^2}\ ,\end{equation}where $b$ is a real
constant. Now by (2.9) we have trivially
\begin{equation}\label{1.16}
{\it
e}^{-U}=(1+a^2u^2)^{-1}+(1+b^2v^2)^{-1}-1=\frac{1-a^2b^2u^2v^2}{(1+a^2u^2)\,(1+b^2v^2)}\
.\end{equation}Also evaluating (2.2), (2.3) and (2.7) on $v=0$ and
on $u=0$ we can obtain the following: on $v=0,\ u>0$ we find that
\begin{equation}\label{1.17}
V_v=2\,a\,b\,u\ ,\qquad \phi _0=b\ ,\end{equation}while on $u=0,\
v>0$ we have
\begin{equation}\label{1.18}
V_u=2\,a\,b\,v\ ,\qquad \phi _2=a\ ,\end{equation}where, as always
a subscript denotes a partial derivative. Using this data the
assumption (2.20) yields
\begin{equation}\label{1.19}
\frac{\phi _2}{\phi _0}=\frac{a\,(1+b^2v^2)}{b\,(1+a^2u^2)}\
,\end{equation} from which we deduce (\ref{1.11}) with
\begin{equation}\label{1.20}
b\,v=\tan\bar v\ ,\qquad a\,u=\tan\bar u\ .\end{equation} Now we
apply the d'Alembert formula (2.23) with $P(\bar u)=0$ and $Q(\bar
u)=2\,\tan\bar u$ to obtain
\begin{equation}\label{1.21}
V=\log\left (\frac{\cos (\bar v-\bar u)}{\cos (\bar u+\bar
v)}\right )=\log\left (\frac{1+a\,b\,u\,v}{1-a\,b\,u\,v}\right )\
.
\end{equation}Now using (2.26), (2.29) and (2.31) we easily see
that
\begin{eqnarray}\label{1.22}
U_v\,\phi _2-V_u\,\phi _0&=&0\ ,\\
U_v\,\phi _0-V_v\,\phi _2&=&0\ .\end{eqnarray}Hence Maxwell's
equations (2.2) and (2.3) reduce to
\begin{equation}\label{1.23}
\frac{\partial\phi _2}{\partial v}=0\ ,\ {\rm and}\ \
\frac{\partial\phi _0}{\partial u}=0\ ,\end{equation}and so using
the initial conditions (2.24) and (2.25) we have, for $u>0\ ,\
v>0$, \begin{equation}\label{1.24} \phi _2=\frac{a}{1+a^2u^2}\
,\qquad \phi _0=\frac{b}{1+b^2v^2}\ .\end{equation}We now
substitute the functions $U,\ V,\ \phi _2\,,\ \phi _0$ given by
(2.26), (2.31) and (\ref{1.24}) respectively into the remaining
field equations (2.5)--(2.8). Equations (2.5) and (2.6) reduce to
\begin{eqnarray}\label{1.25}
M_u&=&\frac{2\,b^2v}{1+b^2v^2}\ ,\\
M_v&=&\frac{2\,a^2u}{1+a^2u^2}\ .\end{eqnarray} Equation (2.7) is
identically satisfied while equation (2.8) becomes
\begin{equation}\label{1.26}
M_{uv}=0\ ,\end{equation}which is clearly consistent with (2.36)
and (2.37). With the boundary conditions (2.24) and (2.25) we
have, for $u>0\ ,\ v>0$,
\begin{equation}\label{1.27}
M=\log (1+a^2u^2)\,(1+b^2v^2)\ .\end{equation} The functions
(2.26), (2.31), (2.35) and (2.39) constitute the Bell--Szekeres
solution of the Einstein--Maxwell vacuum field equations.
Calculation of the Weyl tensor components (2.10)--(2.14) reveals
that they all vanish and so the Bell--Szekeres space--time $u>0\
,\ v>0$ after the collision of the electromagnetic shock waves is
conformally flat.

\setcounter{equation}{0}
\section{Some Einstein--Maxwell Space--Times}\indent
In the Bell--Szekeres example the histories of the wave fronts of
the two families of incoming electromagnetic shock waves have
equations $u={\rm constant}\geq 0\ ,\ v<0$ and $v={\rm
constant}\geq 0\ ,\ u<0$. It is a simple matter to add to these
signals light--like shells and impulsive gravitational waves with
histories $u=0\ ,\ v<0$ and $v=0\ ,\ u<0$. This is done by
modifying the initial conditions (\ref{1.14}) and (\ref{1.15}) to
read (see Appendix A):for $v=0\ ,\ u>0$ we require
\begin{eqnarray}\label{2.1}
{\it e}^{-U}&=&\frac{1-2\,l\,u+(l^2-k^2)\,u^2}{1+a^2u^2}\ ,\\
{\it e}^V&=&\frac{1+(k-l)\,u}{1-(k+l)\,u}\ ,\\
{\it e}^M&=&1+a^2u^2\ ,\\
\phi _2&=&\frac{a}{1+a^2u^2}\ ,\end{eqnarray}and for $u=0\ ,\ v>0$
we require
\begin{eqnarray}\label{2.2}
{\it e}^{-U}&=&\frac{1-2\,p\,v+(p^2-s^2)\,v^2}{1+b^2v^2}\ ,\\
{\it e}^V&=&\frac{1+(s-p)\,v}{1-(s+p)\,v}\ ,\\
{\it e}^M&=&1+a^2u^2\ ,\\
\phi _0&=&\frac{b}{1+b^2v^2}\ .\end{eqnarray}The parameters here
have the following physical associations (in the sense that if any
of the parameters are put equal to zero then that part of the
light--like signal is removed): The parameters $a, b$ label the
incoming electromagnetic shock waves as in section 2. The
parameters $l, p$ label incoming light--like shells while the
parameters $k, s$ label incoming impulsive gravitational waves.
\emph{This collision problem with all parameters non--zero is
unsolved}. Many special cases are of course solved, most
fundamentally the Bell--Szekeres \cite{BS} ($l=p=k=s=0$) case and
the Khan--Penrose \cite{KP} ($a=b=l=p=0$) case. Few cases in which
at least one electromagnetic shock wave is present have been
solved. The solution for a collision involving one electromagnetic
shock wave and two impulsive gravitational waves was derived by
Barrab\`es, Bressange and Hogan \cite{BBH} (see also \cite{BH}).
We have also found the solution for a collision involving one
electromagnetic shock wave labelled by $b$ and two light--like
shells labelled by $l$ and $p$ (thus corresponding to $a=k=s=0$
above). It is given, for $u>0\ ,\ v>0$ by
\begin{eqnarray}\label{2.3}
{\it e}^{-U}&=&(1-l\,u)^2+\frac{(1-p\,v)^2}{1+b^2v^2}-1\ ,\\
{\it
e}^{-M-\frac{1}{2}U}&=&\frac{(1-p\,v)\,(1-l\,u)}{(1+b^2v^2)^{3/2}}\
,\\
\phi _0\,{\it
e}^{-\frac{1}{2}U}&=&\frac{b\,(1-p\,v)}{(1+b^2v^2)^{3/2}}\
,\end{eqnarray}and, in addition $V=0=\phi _2$.

Our objective in this paper is to solve the Einstein--Maxwell
initial value problem with initial data (3.1)--(3.8) and obtain
solutions which include the Bell--Szekeres solution as a special
case. Our strategy to achieve this is to focus on some feature of
the Bell--Szekeres space--time which we can impose on the
collision space--time that we are looking for. Perhaps the
simplest feature of the Bell--Szekeres space--time is its
conformal flatness so we will look for conformally flat solutions
of the Einstein--Maxwell vacuum field equations with the initial
data (3.1)--(3.8). We implement this requirement in a gradual way
which we describe in detail in the next section.

\setcounter{equation}{0}
\section{Generalizations of the Bell--Szekeres Solution}\indent
We begin by searching for necessary conditions for conformal
flatness of the collision space--time. The simplest is obtained by
requiring $\Psi _2$ in (2.12) to vanish at $u=0=v$ (strictly
speaking in the limits $u\rightarrow 0^+$ and $v\rightarrow 0^+$).
With the initial data (3.1)--(3.8) this requirement yields the
following relationship between the parameters:
\begin{equation}\label{3.1}
p\,l=s\,k\ .\end{equation}Another piece of useful information can
be obtained by requiring $\Psi _2$ to vanish \emph{near} $v=0=u$.
To find this we we need to calculate $\phi _0$ and $V_v$ when
$v=0$. The differential equations for these quantities are
obtained by evaluating (2.3) and (2.7) at $v=0$ and using the
initial data. We find that at $v=0\ ,u>0$:
\begin{equation}\label{3.2}
\phi
_0=\frac{p\,a}{k}+\frac{(b\,k-a\,p)}{k\,\sqrt{1-2\,l\,u+(l^2-k^2)\,u^2}}\
,\end{equation}and
\begin{equation}\label{3.3}
V_v=\frac{2\,(b\,k-a\,p)\,a\,u}{k\,\sqrt{1-2\,l\,u+(l^2-k^2)\,u^2}}+\frac{2\,p\,\{
l-(l^2-k^2-a^2)\,u+l\,a^2u^2\}}{k\,(1-2\,l\,u+(l^2-k^2)\,u^2)}\
.\end{equation}At this point, on account of the well--known fact
(2.9), the function $U$ is given for $u>0\ ,\ v>0$, by
\begin{equation}\label{3.4}
{\it
e}^{-U}=\frac{1-2\,l\,u+(l^2-k^2)\,u^2}{1+a^2u^2}+\frac{1-2\,p\,v+(p^2-s^2)\,v^2}{1+b^2v^2}-1\
.
\end{equation}Using these we find that at $v=0$
\begin{equation}\label{3.5}
\Psi
_2=\frac{(b\,k-a\,p)\,a\,u}{(1-2\,l\,u+(l^2-k^2)\,u^2)^{3/2}}\
,\end{equation}from which we conclude that, in addition to
(\ref{3.1}) a further necessary condition for conformal flatness
is
\begin{equation}\label{3.6}
b\,k=a\,p\ .\end{equation}Eq.(4.1) is symmetrical under
interchange of the light--like shell ($l$) and impulsive
gravitational wave ($k$) accompanying the electromagnetic shock
wave ($a$) with the light--like shell ($p$) and impulsive
gravitational wave ($s$) accompanying the electromagnetic shock
wave ($b$) [assuming these shells and gravitational waves exist].
This suggests that (4.6) should have a partner equation which is
obtained from (4.6) under a similar interchange of shell and
gravitational wave. This can be obtained by replacing (4.5) by the
expression for $\Psi _2$ evaluated on $u=0$. More simply, since in
particular we are assuming the $l\,,\ k\,,\ s\,,\ p\neq 0$, if
(4.6) is multiplied by $s$ and (4.1) is used we immediately arrive
at
\begin{equation}\label{3.6'}
a\,s=b\,l\ .\end{equation}

We now have, on account of (3.4) and (4.2) with (4.6) holding,
that when $v=0\ ,\ u>0$,
\begin{equation}\label{3.7}
\phi _0=b\ ,\qquad \phi _2=\frac{a}{1+a^2u^2}\
,\end{equation}while conversely when $u=0\ ,\ v>0$,
\begin{equation}\label{3.8}
\phi _2=a\ ,\qquad \phi _0=\frac{b}{1+b^2v^2}\ .\end{equation}We
now make the assumption (2.20) from which we again arrive at
(2.29) (on account of (\ref{3.7}) and (\ref{3.8})). Introducing
the barred coordinates $\bar u\ ,\ \bar v$ via $b\,v=\tan\bar v$
and $a\,u=\tan\bar u$ as before we have
\begin{equation}\label{3.9}
\bar\phi _0=\bar\phi _2\ .\end{equation}It thus follows that when
$\bar v=0\ ,\ \bar u>0$,
\begin{equation}\label{3.7'}
\bar\phi _0=\bar\phi _2=1\ ,\end{equation}and similarly when $\bar
u=0\ ,\ \bar v>0$. Now the initial data for the wave equation
(2.22) reads
\begin{equation}\label{3.10}
P(\bar u)=\log\left [\frac{a+(k-l)\,\tan\bar u}{a-(k+l)\,\tan\bar
u}\right ]\ ,\end{equation}and (written in a convenient form for
performing the integration in (2.23))
\begin{equation}\label{3.11}
Q(\bar u)=\frac{a\,\tan\bar u+(k+l)}{a-(k+l)\,\tan\bar
u}+\frac{a\,\tan\bar u-(k-l)}{a+(k-l)\,\tan\bar u}\
.\end{equation}The d'Alembert formula (2.23) gives the solution
$V$ in the barred coordinates as
\begin{equation}\label{3.12}
V=\log\left [\frac{a\,\cos (\bar u-\bar v)+(k-l)\,\sin (\bar
u-\bar v)}{a\,\cos (\bar u+\bar v)-(k+l)\,\sin (\bar u+\bar
v)}\right ]\ .\end{equation}

At this point it is useful to define the functions
\begin{eqnarray}\label{3.13}
F_1(\bar u-\bar v)&=&\cos (\bar u-\bar v)+\frac{(k-l)}{a}\,\sin
(\bar u-\bar v)\ ,\\
F_2(\bar u+\bar v)&=&\cos (\bar u+\bar v)-\frac{(k+l)}{a}\,\sin
(\bar u+\bar v)\ .\end{eqnarray} We see that these are wave
functions and also solutions of the unit frequency harmonic
oscillator equation. In terms of them we can write $U$ given by
(4.4) and $V$ by (4.14) in the simple form
\begin{equation}\label{3.14}
{\it e}^{-U}=F_1\,F_2\ ,\qquad {\it e}^V=\frac{F_1}{F_2}\
.\end{equation}From this we see that
\begin{equation}\label{3.14'}
U_{\bar v}=V_{\bar u}\ ,\qquad U_{\bar u}=V_{\bar v}\
,\end{equation} and using these in Maxwell's equations (2.2) and
(2.3) written in the barred coordinates, together with (\ref{3.9})
and the initial data (\ref{3.7'}), we immediately see that for
$\bar u>0\ ,\ \bar v>0$ we have
\begin{equation}\label{3.15}
\bar\phi _0=\bar\phi _2=1\ .\end{equation}Equations (\ref{3.14})
and (\ref{3.15}) are very easy to calculate with. The
Einstein--Maxwell field equations (2.5) and (2.6) in the barred
coordinates easily reduce to
\begin{equation}\label{3.16}
\bar M_{\bar u}=0=\bar M_{\bar v}\ ,\end{equation}and now it
follows that the remaining Einstein--Maxwell field equations (2.7)
and (2.8) are automatically satisfied. With $\bar M$ given in
terms of $M$ by (2.19) the initial data for $\bar M$ read: when
$\bar v=0\ ,\ \bar u>0$, $\bar M=\log (a\,b)$ and when $\bar u=0\
,\ \bar v>0$, $\bar M=\log (a\,b)$. Thus on account of
(\ref{3.16}) we have for $\bar u>0\ ,\ \bar v>0$,
\begin{equation}\label{3.17}
\bar M=\log (a\,b)\ .\end{equation}Finally it is straightforward
to see from (\ref{3.14}) that
\begin{equation}\label{3.18}
V_{\bar u\bar u}-U_{\bar u}\,V_{\bar u}=0=V_{\bar v\bar v}-U_{\bar
v}\,V_{\bar v}\ ,\end{equation}which together with (\ref{3.14'})
help to confirm that the Weyl tensor components (2.10)--(2.14) all
vanish in the barred coordinates for $\bar u>0\ ,\ \bar v>0$ and
so the collision space--time that we have constructed is indeed
conformally flat.

It is interesting to restore the unbarred coordinates $(u, v)$
using $b\,v=\tan\bar v$ and $a\,u=\tan\bar u$. The function $U$ is
given in the coordinates $(u, v)$ by (4.4) which, in the light of
(4.17), can be simplified to read
\begin{equation}\label{3.18'}
{\it
e}^{-U}=\frac{[1+a\,b\,u\,v+(k-l)\,u-(p-s)\,v]\,[1-a\,b\,u\,v-(k+l)\,u
-(p+s)\,v]}{(1+a^2u^2)\,(1+b^2v^2)}\ .\end{equation} By (4.13) the
function $V$ takes the form
\begin{equation}\label{3.19}
V=\log\left
[\frac{1+a\,b\,u\,v+(k-l)\,u-(p-s)\,v}{1-a\,b\,u\,v-(k+l)\,u-(p+s)\,v}
\right ]\ .\end{equation}We have made use of (4.6) and (4.7) to
simplify this expression. Finally the functions $\phi _0\,,\ \phi
_2\,,\ M$ are given by
\begin{equation}\label{3.20}
\phi _0=\frac{b}{1+b^2v^2}\ ,\ \phi _2=\frac{a}{1+a^2u^2}\ ,\
M=\log (1+a^2u^2)\,(1+b^2v^2)\ .\end{equation}The Bell--Szekeres
solution is an allowable special case satisfying the conditions
(4.1), (4.6) and (4.7) with $k=l=p=s=0$ and we now see that
(\ref{3.18'})), (\ref{3.19}) and (\ref{3.20}) agree with (2.26),
(2.31), (2.35) and (2.39) in this case.

\setcounter{equation}{0}
\section{Consequences of Conformal Flatness}\indent
We begin by examining the mathematical consistency with the field
equations of the assumption of conformal flatness in the region of
the space--time after the collision. From (2.10)--(2.14) we have
the necessary and sufficient conditions for conformal flatness:
\begin{eqnarray}\label{5.1}
V_{vv}&=&U_v\,V_v-M_v\,V_v\ ,\\
U_u\,U_v&=&V_u\,V_v\ ,\\
V_{uu}&=&U_u\,V_u-M_u\,V_u\ .\end{eqnarray}Differentiating (5.2)
with respect to $v$ and using the field equations (given in
(2.2)--(2.8)) we find that provided $\phi _0\neq 0$ we must have
\begin{equation}\label{5.2}
U_u\,\phi _0=V_v\,\phi _2\ ,\end{equation}and thus (2.3) implies
\begin{equation}\label{5.3}
\phi _0=\phi _0(v)\ .\end{equation} Similarly differentiating
(5.2) with respect to $u$ we find that if $\phi _2\neq 0$ then
\begin{equation}\label{5.4}
U_v\,\phi _2=V_u\,\phi _0\ ,\end{equation} and so the Maxwell
equation (2.2) implies
\begin{equation}\label{5.5}
\phi _2=\phi _2(u)\ .\end{equation}From (5.5) and (5.7) we see in
particular that we have the separation of variables (2.20) as a
consequence of conformal flatness. Next differentiating (5.1) with
respect to $u$, using the field equations, and assuming that $\phi
_2\neq 0$ we arrive at
\begin{equation}\label{5.6}
\frac{d\phi _0}{dv}=-M_v\,\phi _0\ ,\end{equation}while
differentiating (5.2) with respect to $v$ we find that if $\phi
_0\ne 0$ then
\begin{equation}\label{5.7}
\frac{d\phi _2}{du}=-M_u\,\phi _2\ .\end{equation}On account of
the field equation (2.8) we see from (5.2) that
\begin{equation}\label{5.8}
M=A(u)+B(v)\ ,\end{equation}where the functions $A$ and $B$ are
arbitrary. Using this in (5.8) and (5.9) we obtain
\begin{equation}\label{5.9}
\phi _0=c_0\,{\it e}^{-B(v)}\ ,\qquad \phi _2=c_2\,{\it
e}^{-A(u)}\ ,\end{equation}where $c_0$ and $c_2$ are constants. We
solve (5.4) and (5.6) for $V_v$ and $V_u$ to obtain
\begin{equation}\label{5.10}
V_v=\frac{\phi _0}{\phi _2}\,U_u\ ,\qquad V_u=\frac{\phi _2}{\phi
_0}\,U_v\ .\end{equation}It is now clear that with $U$ given by
(2.9) the equations (5.1)--(5.3) are satisfied. Using (5.8) and
(5.9) the integrability conditions for (5.12) read
\begin{equation}\label{5.11}
\phi _2^2(U_{vv}+U_v\,M_v)=\phi _0^2(U_{uu}+U_u\,M_u)\
.\end{equation} Substituting from the field equations (2.5) and
(2.6) this becomes
\begin{equation}\label{5.12}
\phi _2^2(U_v^2+V_v^2)=\phi _0^2(U_u^2+V_u^2)\ ,\end{equation}and
this equation is satisfied on account of (5.12).

To examine the status of (2.7) we use the first of (5.12) along
with (5.9) to obtain
\begin{equation}\label{5.13}
2\,V_{vu}=2\,\frac{\phi _0}{\phi _2}\,(U_{uu}+U_u\,M_u)\
.\end{equation}By (2.5) this reads
\begin{eqnarray}\label{5.14}
2\,V_{vu}&=&\frac{\phi _0}{\phi _2}\,(U_u^2+V_u^2+4\,\phi _2^2)\
,\nonumber\\
&=&V_v\,U_u+V_u\,U_v+4\,\phi _0\,\phi _2\ ,\end{eqnarray}using
(5.12) again. Thus we see that (2.7) is satisfied. From now on we
therefore concentrate attention on (2.5) and (2.6).

In view of (2.19) we can take advantage of (5.11) to introduce
barred coordinates $\bar u(u)\,,\ \bar v(v)$ via the differential
equations
\begin{equation}\label{5.15}
\frac{d\bar u}{du}=c_2\,{\it e}^{-A(u)}\ ,\qquad \frac{d\bar
v}{dv}=c_0\,{\it e}^{-B(v)}\ .\end{equation} This has the
immediate effect of having
\begin{equation}\label{5.16}
\bar\phi _0=\bar\phi _2=1\ ,\end{equation}and, on account of
(5.10), of also having
\begin{equation}\label{5.17}
\bar M=\log (c_0\,c_2)\ .\end{equation}The field equations
(2.2)--(2.8) remain invariant in form under the transformation to
the barred coordinates and hence (2.4) gives, in the barred
system, \begin{equation}\label{5.18} {\it e}^{-U}={\cal F}(\bar
u)+{\cal G}(\bar v)\ ,\end{equation}where ${\cal F}\,,\ {\cal G}$
are arbitrary while the remaining field equations of interest,
(2.5) and (2.6), simplify in the barred system to
\begin{eqnarray}\label{5.18}
2\,U_{\bar u\bar u}&=&U_{\bar u}^2+V_{\bar u}^2+4\ ,\\
2\,U_{\bar v\bar v}&=&U_{\bar v}^2+V_{\bar v}^2+4\ .\end{eqnarray}
We also have (5.12), which in the barred system reduces to
\begin{equation}\label{5.19}
U_{\bar u}=V_{\bar v}\ ,\qquad U_{\bar v}=V_{\bar u}\
.\end{equation}Thus we will solve for the initial data functions
${\cal F}\,,\ {\cal G}$ using (5.21) and (5.22) in the form
\begin{equation}\label{5.20}
2\,U_{\bar u\bar u}=U_{\bar u}^2+U_{\bar v}^2+4=2\,U_{\bar v\bar
v}\ .\end{equation}The two equations we obtain for ${\cal F}\,,\
{\cal G}$ can be written in the form
\begin{equation}\label{5.21}
\frac{d^2{\cal F}}{d\bar u^2}+4\,{\cal F}=-\left (\frac{d^2{\cal
G}}{d\bar v^2}+4\,{\cal G}\right )\ ,\end{equation}and
\begin{equation}\label{5.22}
({\cal F}+{\cal G})\,\left (\frac{d^2{\cal G}}{d\bar
v^2}-\frac{d^2{\cal F}}{d\bar u^2}\right )=\left (\frac{d{\cal
G}}{d\bar v}\right )^2-\left (\frac{d{\cal F}}{d\bar u}\right )^2\
.\end{equation}The solutions of (5.25) are
\begin{eqnarray}\label{5.23}
{\cal F}&=&\alpha _0\,\sin 2\,\bar u+\beta _0\,\cos 2\,\bar u+k_0\
,\\{\cal G}&=&\gamma _0\,\sin 2\,\bar v+\delta _0\,\cos 2\,\bar
v-k_0\ ,\end{eqnarray}where $k_0$ is a separation constant and
$\alpha _0, \beta _0,\ \gamma _0, \delta _0$ are constants of
integration. Substitution into (5.26) yields
\begin{equation}\label{5.24}
\alpha _0^2+\beta _0^2=\gamma _0^2+\delta _0^2\
.\end{equation}Writing $\alpha _0+i\beta _0=R_0\,{\it e}^{2i\xi
_0}$ and $\gamma _0+i\delta _0=R_0\,{\it e}^{2i\eta _0}$ we can
write (5.20) as
\begin{equation}\label{5.25}
{\it e}^{-U}=C\,\{\cos (\bar u+\bar v)+\cot (\xi _0+\eta _0)\,\sin
(\bar u+\bar v)\}\,\{\cos (\bar u-\bar v)-\tan (\xi _0-\eta
_0)\,\sin (\bar u-\bar v)\}\ ,\end{equation}with
\begin{equation}\label{5.26}
C=2\,R_0\,\sin (\xi _0+\eta _0)\,\cos (\xi _0-\eta _0)={\rm
constant}\ .\end{equation}Using this in (5.23) we solve for $V$ to
obtain
\begin{equation}\label{5.27}
{\it e}^V=W_0\,\left (\frac{\cos (\bar u-\bar v)-\tan (\xi _0-\eta
_0)\,\sin (\bar u-\bar v)}{\cos (\bar u+\bar v)+\cot (\xi _0+\eta
_0)\,\sin (\bar u+\bar v)}\right )\ ,\end{equation}where $W_0$ is
a constant of integration. For substitution into the line--element
(2.1) we now have the functions
\begin{eqnarray}\label{5.28}
{\it e}^{-U+V}&=&C_1\,\{\cos (\bar u-\bar v)-\tan (\xi _0-\eta
_0)\,\sin
(\bar u-\bar v)\}^2\ ,\\
{\it e}^{-U-V}&=&C_2\,\{\cos (\bar u+\bar v)+\cot (\xi _0+\eta
_0)\,\sin (\bar u+\bar v)\}^2\ ,\end{eqnarray} where
\begin{eqnarray}\label{5.29}
C_1&=&2\,W_0\,R_0\,\sin (\xi _0+\eta _0)\,\cos (\xi _0-\eta _0)\
,\\
C_2&=&2\,\frac{R_0}{W_0}\,\sin (\xi _0+\eta _0)\,\cos (\xi _0-\eta
_0)\ .\end{eqnarray}These constants can be absorbed by a
re-scaling of the coordinates $x, y$ which is equivalent to
putting, without loss of generality, $C=W_0=1$ in (5.30) and
(5.32). Comparing now the expressions for $\bar M, \bar\phi _0,\
\bar\phi _2,\ U,\ V$ which we have derived here in (5.18), (5.19),
(5.30) (with $C=1$) and (5.32) (with $W_0=1$) with the
corresponding expressions (4.19), (4.21) and (4.17), obtained as
the solution of the collision problem in section 4, we see that
they are identical provided the constants $c_0, c_2, \xi _0,\ \eta
_0$ in this section are related to the constants $a, b, k, l$ in
section 4 by
\begin{equation}\label{5.30}
c_0\,c_2=a\,b\ ,\ \ \cot (\xi _0+\eta _0)=-\frac{(k+l)}{a}\ ,\ \
\tan (\xi _0-\eta _0)=-\frac{(k-l)}{a}\ .\end{equation}We have
thus established that if the space--time region after the
collision of the light--like signals is a solution of the vacuum
Einstein--Maxwell field equations and if, in addition, it is
conformally flat, then the colliding light--like signals have to
be those related combinations of shells, impulsive gravitational
waves and electromagnetic shock waves described in section 4.

 \setcounter{equation}{0}
\section{Discussion}\indent
We have focussed on the conformal flatness property of the
Bell--Szekeres solution and carried that into the generalization
derived in section 4. There is however a further property of the
Bell--Szekeres solution which is inherited by the generalization.
To see this we write the line--element of the solution in section
4 in the barred coordinates. It is given by
\begin{equation}\label{4.1}
ds^2=-F_1^2dx^2-F_2^2dy^2+2\,(a\,b)^{-1}d\bar u\,d\bar v\
,\end{equation}with $F_1$ and $F_2$ given by (4.15) and (4.16).
Introducing coordinates $\xi =\bar u -\bar v$ and $\eta =\bar u
+\bar v$ we have \begin{equation}\label{4.2} ds^2=ds_1^2+ds_2^2\
,\end{equation}with
\begin{eqnarray}\label{4.2}
ds_1^2&=&-F_1^2(\xi )\,dx^2-\frac{1}{2\,a\,b}\,d\xi ^2\ ,\\
ds_2^2&=&-F_2^2(\eta )\,dy^2+\frac{1}{2\,a\,b}\,d\eta ^2\
.\end{eqnarray}Hence we see that the collision space--time has
decomposed into the Cartesian product of a pair of
two--dimensional space--times having line--elements (6.3) and
(6.4). The Gaussian curvature of the 2--space with line--element
(6.3) is $K_1=\mp 2\,a\,b$ while the Gaussian curvature of the
2--space with line element (6.4) is $K_2=\pm 2\,a\,b$ - in each
case we have here a 2--space of constant Gaussian curvature of
opposite sign. This is a property that the collision space--times
derived in section 4 share with the Bell--Szekeres space--time. It
is well--known that the Bell--Szekeres space--time coincides with
the Bertotti--Robinson \cite{Ber}\cite{R} space--time which was
originally identified as the four dimensional Einstein--Maxwell
vacuum space--time having this topological property. The solution
given in section 4 represents only \emph{a portion} of the
Bertotti--Robinson space--time because, depending upon the sign of
the product $a\,b$, either the $x$ coordinate is periodic with
period $2\,\pi\,a/\sqrt{a^2+(k-l)^2}<2\,\pi$ or the $y$ coordinate
is periodic with period $2\,\pi\,a/\sqrt{a^2+(k+l)^2}<2\,\pi$. In
both cases the period of the coordinate is $2\,\pi$ in the
Bell--Szekeres special case.

\appendix
\section{In-Coming Signal} \setcounter{equation}{0}
The line--element of the space--time containing the history of the
in--coming signal can be written in the form
\begin{eqnarray}\label{A1}
ds^2&=&-\left (\cos a\,u'_++\frac{(k-l)}{a}\,\sin a\,u'_+\right
)^2dx^2-\left (\cos a\,u'_+-\frac{(k+l)}{a}\,\sin a\,u'_+\right
)^2dy^2\nonumber\\&&+2\,du'\,dv\ .\end{eqnarray} Here
$u'_+=u'\,\vartheta (u')$ where $\vartheta (u')$ is the Heaviside
step function which is equal to zero if $u'<0$ and equal to unity
if $u'>0$. Direct calculation of the Ricci tensor on the half null
tetrad given via the 1--forms
\begin{eqnarray}\label{A2}
\theta ^1&=&\left (\cos a\,u'_++\frac{(k-l)}{a}\,\sin
a\,u'_+\right )\,dx=-\theta
_1\ ,\\
\theta ^2&=&\left (\cos a\,u'_+-\frac{(k+l)}{a}\,\sin
a\,u'_+\right )\,dy=-\theta
_2\ ,\\
\theta ^3&=&dv=\theta _4\ ,\\
\theta ^4&=&du'=\theta _3\ ,\end{eqnarray}results in $R_{ab}\equiv
0$ except for
\begin{equation}\label{A3}
R_{44}=-2\,l\,\delta (u')-2\,a^2\vartheta (u')\
,\end{equation}where $\delta (u')$ is the Dirac delta function.
The $a^2$--term in (\ref{A3}) is due to a vacuum electromagnetic
field. Taking the potential 1--form to be
\begin{equation}\label{A4}
A=\left (\sin a\,u'_+-\frac{(k-l)}{a}\,\cos a\,u'_+\right )\,dx\
,\end{equation} we obtain the field $F$ and its dual ${}^*F$ given
respectively by
\begin{equation}\label{A5}
F=-a\,\vartheta (u')\,\theta ^2\wedge\theta ^4\ ,\qquad {\rm
and}\qquad {}^*F=-a\,\vartheta (u')\,\theta ^1\wedge\theta ^4\
.\end{equation}It follows trivially that Maxwell's vacuum field
equations are satisfied by $F$. Calculation of the electromagnetic
energy--momentum tensor components $E_{ab}$ on the tetrad reveals
that all components vanish except
\begin{equation}\label{A6}
E_{44}=-a^2\,\vartheta (u')\ .\end{equation}Thus (\ref{A3}) may be
re--written in the form
\begin{equation}\label{A7}
R_{44}-2\,E_{44}=-2\,l\,\delta (u')\ .\end{equation}Thus the
Einstein--Maxwell field equations with a light--like shell source
(provided $l\neq 0$) are satisfied by the metric tensor given via
the line--element (\ref{A1}). The only non--vanishing Weyl tensor
component in Newman--Penrose notation is
\begin{equation}\label{A8}
\Psi _4=-R_{1414}+\frac{1}{2}\,R_{44}=-k\,\delta (u')\
,\end{equation} indicating that the signal is accompanied by an
impulsive gravitational wave provided $k\neq 0$. This Weyl tensor
is type N in the Petrov classification with $\partial /\partial v$
as degenerate principal null direction. Hence the key equations
for physically interpreting the signal are (\ref{A6})--(\ref{A8}).
When $u'>0$ the transformation $a\,u=\tan a\,u'$ applied to
(\ref{A1}) yields the initial data quoted in (3.1)--(3.4).

\end{document}